\documentclass{iaus}
\usepackage{graphicx}

\title[~~Opacity of M31 disk] 
{Dust properties of nearby disks: M\,31 case}

\author[P. Nedialkov, A. Valcheva, V. Ivanov \& L. Vanzi]   
{Petko L. Nedialkov$^1$, Antoniya T. Valcheva$^2$, \\ Valentin D.
Ivanov$^3$ \& Leonardo Vanzi$^4$}

\affiliation{$^1$Astronomy Department, University of Sofia, \\
5 J. Bourchier blvd., Sofia 1164, Bulgaria\\ email: {\tt japet@phys.uni-sofia.bg} \\[\affilskip]
$^2$Institute of Astronomy,\\ 72 Tsarigradsko Chaussee blvd.,
Sofia 1784, Bulgaria \\email: {\tt avalcheva@astro.bas.bg}\\[\affilskip]
$^3$European Southern Observatory, \\ 3107 Alonso de Cordova Ave.,
Casilla 19, Santiago 19001, Chile \\email: {\tt vivanov@eso.org}\\[\affilskip]
$^4$Department of Astronomy and Astrophysics, Pontificia Universidad\\ Cat\'olica
de Chile, Casilla 306, Santiago 22, Chile \\email: {\tt lvanzi@eso.org}}

\pubyear{2008}
\volume{254}  
\pagerange{1--7}
 \jname{The Galaxy Disk in Cosmological
Context}
\editors{J. Andersen, J. Bland-Hawthorn \& B. Nordstr\"om, eds.}
\begin{document}

\maketitle

\begin{abstract}
Several properties of the M\,31 disk, namely: opacity, extinction
law and gas-to-dust ratio are studied by means of optical and near-infrared 
photometry of ten globular clusters and galaxies seen
through the disk. The individual extinctions of these objects were
estimated with respect to several sets of theoretical spectral
energy distributions for globulars and galaxies. Seven targets
are consistent with reddened globulars, two - with starburst
galaxies and one - with an elliptical. The extinction estimates
agree with semi-transparent disk ($\tau_V\lesssim1$) in the
inter-arm regions. The total-to-selective extinction ratio in
those regions 2.75$\pm$0.1 is lower on average than the typical
Galactic value of $R_V$=3.1. We also obtained a gas-to-dust ratio,
similar to that in the the Milky way. It shows no correlation with
the distance from the M31 center.

\end{abstract}

\firstsection 

\section[]{Introduction}
The dust in galaxies attenuates the light of background
extragalactic sources. Recent studies (\cite[Holwerda et al.
2005]{Holwerda_etal05}) of morphologically representative samples
of spirals shows that dust opacity of a disk arises from two
distinct components: {\it(i)} optically thicker ($\tau_V=0.8 \div
8^m$) and radially dependent component, associated with the spiral
arms, and {\it(ii)} relatively more constant optically thinner
disk ($\tau_V\sim1^m$), dominating the inter-arm regions and the
outskirts of the disk.

The nearby giant spiral galaxy M\,31 is well suited for
comprehensive studies of the interplay between the stars and the
ISM. The radial distribution of the opacity in M\,31 spiral arms,
based on individual estimates towards OB stars, shows that the
opacity exponentially decreases away from the bulge
(\cite[Nedilakov \& Ivanov 1998]{NedilakovIvanov}). However, a
study of 41 globular clusters in M\,31 indicates the absence of
radial extinction gradient with the galactocentric distance
(\cite[Savcheva \& Tassev 2002]{SavchevaTassev02}).

Measuring the color excesses of objects behind the disks is an
alternative method to constrain the disk opacity. It was applied
to M\,31 by \cite[Cuillandre et al. (2001)]{Cuillandre_etal01}
who used background galaxies. They concluded that the M31 disk is
semi-transparent for distances larger than $R_{25}$.

Here we complement this work, presenting opacity estimates for
galactocentric distance smaller than $R_{25}$, derived from the
comparison of apparent colors of background globulars and
ellipticals with models.

\section{Observations and data reduction}
Our sample includes 21 background galaxy candidates located well
within the standard M\,31 radius $R_{25}$. They were selected from
a number of heterogeneous sources: visual inspection of DSS and
the NOAO archive photographic plates, dropouts from M\,31 globular
cluster searches (\cite[Battistini et al. 1980]{Battistini_etal80}).
Our original intention was to base the study on background
ellipticals only but five of our targets were recently identified
as globulars, and their [Fe/H] and v$_r$ became readily available
from \cite[Galleti et al. (2006)]{Galleti_etal06}.

\begin{figure}[t]
\begin{center}
  \includegraphics[width=3.25in]{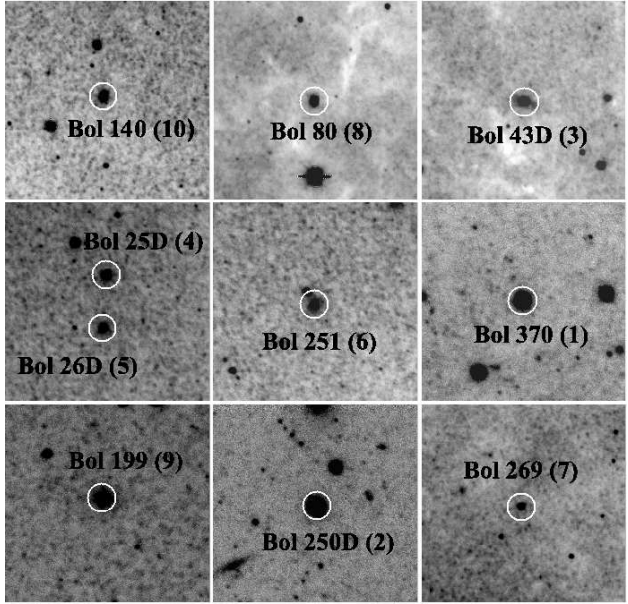}
\caption{$V$-band images of our targets from the Local Group
Survey (\cite[Massey et al. 2006]{Massey_etal06}). The field of
view is $30^{\prime\prime}\times\,30^{\prime\prime}$. North is
up, and East is to the left. The white circles show the
photometric extraction apertures. The numbering corresponds to
Table\,\ref{tab1}.}
\label{fig1}
\end{center}
\end{figure}

We obtained $HK$ imaging in Dec 1996 and Jan 1997 with ARNICA
(\cite[Lisi et al. 1993]{Lisi_etal93}) at 1.8m Vatican Advanced
Technology Telescope on Mt. Graham. The instrument is equipped
with a NICMOS\,3 (256 $\times$ 256 pixels) detector array, with
scale of 0.505 arcsec/pixel. The data reduction includes the
typical steps for infrared imaging: ``sky'' removal, flat-fielding,
alignment and combination of individual images, separately for
every filter and field.
Ten of the targets (Table\,\ref{tab1}, Fig.\,\ref{fig1}) were
identified on the $UBVRI$ images from the Local Group Survey
(\cite[Massey et al. 2006]{Massey_etal06}), obtained with the
KPNO Mosaic Camera at 4m Mayall telescope.

\begin{table}[h]
\begin{center}
\caption{Sub-set of our original target list, with $UBVRIHK$
photometry.} \label{tab1} {\scriptsize
\begin{tabular}{|l|c|c|c|c|c|c|c|}\hline
{\bf No.} & {\bf 2MASX/2MASS } & {\bf Other} & {\bf Object$^1$} & {\bf v$_r$}          & {\bf [Fe/H]} & {\bf r$_{gc}$} & {\bf N(HI+2H$_2$)$^2$} \\
          & {\bf name}         & {\bf name}  & {\bf Type}       & {\bf [km\,s$^{-1}$]} &              & {\bf [arcmin]} & {\bf [$\times$10$^{20}$\,at.\,cm$^{-2}$]} \\ \hline
 1. & 2MASX\,J00451437+4157405 & Bol\,370  & 1 & $-$347 & $-$1.80 & 52.2 &  1.895 \\ \hline
 2. & 2MASX\,J00444399+4207298 & Bol\,250D & 1 & $-$442 &   -     & 84.2 &  9.028 \\ \hline
 3. & 2MASS\,J00420658+4118062 & Bol\,43D  & 1 & $-$344 & $-$1.35 & 30.8 & 12.836 \\ \hline
 4. & 2MASS\,J00413428+4101059 & Bol\,25D  & 1 & $-$479 &   -     & 20.8 &  0.668 \\ \hline
 5. & 2MASS\,J00413436+4100497 & Bol\,26D  & 1 & $-$465 & $-$1.15 & 20.8 &  2.170 \\ \hline
 6. & 2MASS\,J00413660+4100182 & Bol\,251  & 2 & -      &   -     & 20.6 &  1.350 \\ \hline
 7. & 2MASS\,J00430737+4127329 & Bol\,269  & 2 & -      &   -     & 19.6 & 12.384 \\ \hline
 8. & 2MASS\,J00421236+4119008 & Bol\,80   & 2 & -      &   -     & 29.3 & 15.248 \\ \hline
 9. & 2MASX\,J00410351+4029529 & Bol\,199  & 2 & -      & $-$1.59 & 79.1 & 16.120 \\ \hline
10. & 2MASS\,J00425875+4108527 & Bol\,140  & 3 & $-$413 & $-$0.88 & 31.7 &  6.654 \\ \hline
\end{tabular}
}
\end{center}
\vspace{1mm}
\scriptsize{
{\it Notes:}\\
$^1$ Following Galleti et al. (2006):
1 - confirmed globular clusters (GC),
2 - GC candidates,
3 - uncertain objects \\
$^2$Total hydrogen column density based on pencil beam estimates
of CO(1$\rightarrow$0) intensity (\cite[Nieten et al.
2006]{Nieten_etal06}), converted to molecular hydrogen column
density using a constant $X_{\rm CO}$ conversion factor
(\cite[Strong \& Mattox 1996]{StrongMattox96}) and
$\lambda$21\,cm emission from the Westerbork map (\cite[Brinks \&
Shane 1984)]{BrinksShane84}.}
\end{table}

\begin{figure}[h]
\begin{center}
  \includegraphics[width=4.25in]{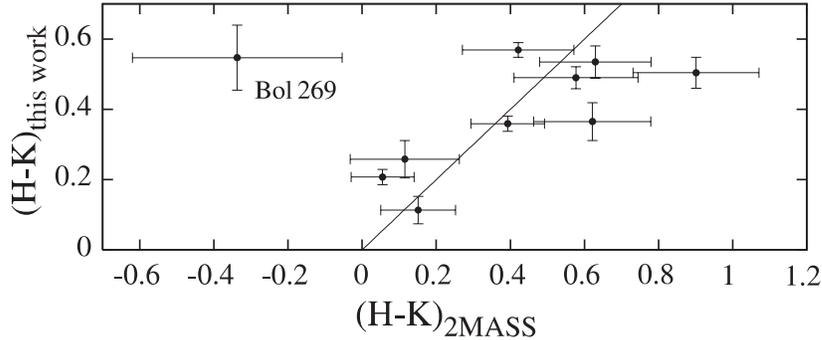}
\caption{Comparison between the colors derived in this work and
the available 2MASS colors. Note that the photometry of Bol\,269
(target No.\,7 in our list) was flagged as suspect in the 2MASS
Point Source Catalog.}
   \label{fig2}
\end{center}
\end{figure}

Clouds were present during most of the observations, forcing us to
use the 2MASS Point Source Catalog (\cite[Cutri et al.
1993]{Cutri2003}) stars for the photometric calibration (typically
using 4--10 common stars per field). No color dependence was
found, and the r.m.s. of the derived zero-points was
$\sim$0.05\,mag for both bands. The typical seeing of the
optical images ($\sim$$1.0^{\prime\prime}$) matches well that of
the near-infrared data set ($\sim$$1.5^{\prime\prime}$), allowing
us to perform simple aperture photometry with $4^{\prime\prime}$
radius. We used the standard IRAF\footnote {IRAF is the Image
Analysis and Reduction Facility made available to the astronomical
community by the National Optical Astronomy Observatories, which
are operated by AURA, Inc., under contract with the U.S. National
Science Foundation.} tasks. The zero points of the optical data
are based on stars in common with the catalog of \cite[Massey et
al. (2006)]{Massey_etal06}. The $V$-band magnitudes and the
observed colors together with their errors are listed in
Table\,\ref{tab2}. The majority of the infrared colors shows
excellent agreement with the available 2MASS colors (see Fig.\,\ref{fig2}).

\begin{table}[h]
\begin{center}
\caption{$UBVRIHK$ photometry of the objects listed in
Table\,\ref{tab1}. Photometric systems: $UBV$ -- Johnson,
$RI$ - Cousins, $HK$ - \cite[Bessell \& Brett
(1988)]{BessellBrettl06}. The uncertainties include both
the zero-point errors and the statistical errors of the
individual measurements.}
 \label{tab2}
 {\scriptsize
\begin{tabular}{|l|c|c|c|c|c|c|c|c|c|c|c|c|c|c|}\hline
{\bf No.} & {\bf $V$} & {\bf $\sigma_V$} & {\bf $U-K$} & {\bf
$\sigma_{U-K}$} & {\bf $U-I$} & {\bf $\sigma_{U-I}$}& {\bf $U-R$}
& {\bf $\sigma_{U-R}$}& {\bf $B-K$} & {\bf $\sigma_{B-K}$} & {\bf
$V-K$} & {\bf $\sigma_{V-K}$} & {\bf $U-H$} & {\bf $\sigma_{U-H}$}
\\ \hline
 1. & 16.18  & 0.08 &  3.85 &  0.30 &  2.23 & 0.31 &  1.84 &  0.32 &  3.02 &  0.23 &  2.53 &  0.10 &  3.67 &  0.30 \\ \hline
 2. & 17.60  & 0.02 &  6.19 &  0.09 &  3.51 & 0.06 &  2.69 &  0.06 &  5.11 &  0.09 &  4.25 &  0.08 &  5.65 &  0.06 \\ \hline
 3. & 17.15  & 0.02 &  6.46 &  0.05 &  3.90 & 0.06 &  3.08 &  0.05 &  5.16 &  0.07 &  4.21 &  0.05 &  6.13 &  0.06 \\ \hline
 4. & 18.23  & 0.03 &  6.25 &  0.07 &  3.49 & 0.10 &  2.80 &  0.05 &  5.34 &  0.10 &  4.19 &  0.06 &  5.74 &  0.07 \\ \hline
 5. & 18.23  & 0.03 &  5.72 &  0.07 &  3.43 & 0.10 &  2.70 &  0.05 &  4.85 &  0.10 &  3.81 &  0.07 &  5.39 &  0.07 \\ \hline
 6. & 17.77  & 0.03 &  5.82 &  0.07 &  3.46 & 0.10 &  2.79 &  0.04 &  4.65 &  0.09 &  3.69 &  0.06 &  5.35 &  0.06 \\ \hline
 7. & 18.49  & 0.03 &  5.83 &  0.09 &  3.51 & 0.07 &  2.96 &  0.06 &  4.68 &  0.11 &  3.72 &  0.09 &  5.31 &  0.09 \\ \hline
 8. & 17.14  & 0.02 &  5.10 &  0.06 &  3.28 & 0.06 &  2.59 &  0.04 &  3.98 &  0.08 &  3.25 &  0.05 &  5.01 &  0.06 \\ \hline
 9. & 18.26  & 0.02 &  6.36 &  0.07 &  3.76 & 0.07 &  2.95 &  0.06 &  5.20 &  0.08 &  4.27 &  0.04 &  5.90 &  0.08 \\ \hline
10. & 17.37  & 0.02 &  4.64 &  0.07 &  2.38 & 0.12 &  2.19 &  0.04
& 3.59 & 0.09 &  3.08  & 0.07  & 4.43 &  0.07 \\ \hline
\end{tabular}
}
\end{center}
\end{table}

\section{Dust properties of M\,31 disk from $\chi^2$-test minimization}

The intrinsic near-infrared colors of ellipticals are nearly
identical: ($H$$-$$K$)$_0$$\sim$0.22\,mag, as demonstrated by
\cite[Persson et al. (1979)]{Persson_etal79}. Assuming all the
targets belong to that Hubble type, the opacity of the disk
that lay between them and the observer can easily be calculated,
taking into account the internal Milky Way extinction, i.e.
from the work of \cite[Schlegel et. al (1998)]{Schlegel_etal98}.
However, the evident contamination of the sample by globular
clusters requires also to consider for each object the
possibility that it is a M\,31 globular, with the typical
globular cluster colors. Furthermore, a cluster may be located
in front of the M\,31 disk, adding extra degree of complication
to the analysis.

To address these issues we developed a multicolor $\chi^2$
minimization technique to derive simultaneously the disk opacity
and number of other parameters: gas-to-dust ratio, extinction
law and last but not least -- the nature of the object
(elliptical galaxy or globular cluster). It allows also to fix
some of these parameters, while still varying the rest of them.
The intrinsic colors of globulars were adopted from the model
of \cite[Kurth et al. (1999)]{Kurth_etal99} and for the
ellipticals -- from \cite[Bicker et al. (2004)]{Bicker_etal04}.

The results from the test are presented in Table\,\ref{tab3}.
The free parameters in the case of globular clusters (left side
of the Table\,\ref{tab3}) are: age, abundance $Z$, $R_V$, and
in the case of the elliptical galaxy models (right side of the
Table\,\ref{tab3}) they are: redshift $z$, $R_V$, and $A_V$.
We created a multi-dimensional grid, with steps of 0.01 along
all axes and calculated the $\chi^2$ for every grid node.

\begin{table}[t]
\begin{center}
\caption{Summary of the $\chi^2$ minimization.
The matches of the apparent colors to the globular cluster models of
\cite[Kurth et al. (1999)]{Kurth_etal99} is given in the left and to the
intrinsic colors of ellipticals predicted by the GALEV models of
\cite[Bicker et al. (2004)]{Bicker_etal04} is given on the right. The
numbers of the targets as the same as in Table\,1. The table also reports
if the $A_V$ corresponding to the minimum $\chi^2$ agrees (within the
errors) with the total gas density derived from the combined $H I$ map of
\cite[Brinks \& Shane (1984)]{BrinksShane84} and the CO(1$\rightarrow$0)
map of \cite[Nieten et al. (2006)]{Nieten_etal06}. The asterisk indicates
a fixed parameter.}
 \label{tab3}
 {\scriptsize
\begin{tabular}{|l||c|l|l|c|r|c||c|l|c|r|c||c|}\hline

{\bf No.} & \multicolumn{5}{c}{{\bf \,\,\,\,\,\,\,Globular Cluster Model Fit}} & & \multicolumn{4}{c}{{\bf \,\,\,\,Elliptical Galaxy Model Fit}} & & {\bf Derived} \\
          & \multicolumn{5}{c}{{\bf \,\,\,\,\,\,\,\,\,\,\,\,\,\,Parameters}} & & \multicolumn{4}{c}{{\bf \,\,\,\,\,\,\,\,\,\,\,\,\,\,Parameters}} & & {\bf Type} \\ \hline

          & {\bf Age}  & {\bf Abun-}       & {\bf $R_V$} & {\bf $A_V$} & {\bf $\chi_{min}^2$} & {\bf dust} & {\bf Red-} & {\bf $R_V$} & {\bf $A_V$} & {\bf $\chi_{min}^2$} & {\bf dust} & \\
          & {\bf [yr]} & {\bf dance}       &             & {\bf [mag]} &                      & {\bf vs.}  & {\bf shift}&             & {\bf [mag]} &                      & {\bf vs.}  & \\
          &            & \,\,\,\,\,\,\,{\bf $Z$} &             &             &                      & {\bf gas?} & {\bf $z$}  &             &             &                      & {\bf gas?} & \\ \hline

 1. & 0.9$\times$10$^9$    & 0.0001  & 3.10* & 1.81 &  1.288 & no  & 0.000 & 3.10* & 0.00 &  39.950 & no  & Globular \\
    & 0.9$\times$10$^9$    & 0.0003* & 3.10* & 1.46 &  2.510 & no  & 0.000 & 6.00  & 0.00 &  39.950 & no  & \\
    & 0.6$\times$10$^9$    & 0.0003* & 2.43  & 1.45 &  1.534 & no  &       &       &      &         &     & \\ \hline

 2. & 0.4$\times$10$^9$    & 0.0500  & 3.10* & 2.41 &  2.400 & no  & 0.075 & 3.10* & 0.98 &   7.562 & yes & Globular \\
    &                      &         &       &      &        &     & 0.069 & 3.42  & 1.05 &   6.900 & yes & \\ \hline

 3. & 0.8$\times$10$^9$    & 0.0500  & 3.10* & 0.50 &  3.406 & yes & 0.026 & 3.10* & 1.41 &  45.111 & yes & Globular\\
    & 3.0$\times$10$^9$    & 0.0009* & 3.10* & 2.60 & 19.558 & no  & 0.021 & 2.15  & 1.18 &  10.404 & yes & \\
    & 2.0$\times$10$^9$    & 0.0009* & 2.77  & 2.54 &  8.552 & no  &       &       &      &         &     & \\ \hline

 4. & 3.0$\times$10$^9$    & 0.0500  & 3.10* & 0.86 &  2.152 & no  & 0.072 & 3.10* & 1.06 &   7.180 & no  & Globular\\
    &                      &         &       &      &        &     & 0.080 & 2.75  & 0.97 &   5.692 & no  & \\ \hline

 5. & 8.0$\times$10$^9$    & 0.0400  & 3.10* & 0.21 &  0.870 & yes & 0.046 & 3.10* & 0.86 &  24.117 & no  & Globular\\
    & 1.1$\times$10$^{10}$ & 0.0014* & 3.10* & 1.73 &  7.522 & no  & 0.045 & 1.77  & 0.62 &   3.649 & no  & \\
    & 2.0$\times$10$^9$    & 0.0014* & 2.75  & 2.00 &  2.786 & no  &       &       &      &         &     & \\ \hline

 6. & 1.4$\times$10$^{10}$ & 0.0400  & 3.10* & 0.01 &  1.707 & no  & 0.061 & 3.10* & 0.83 &  69.838 & no  & Globular \\
    &                      &         &       &      &        &     & 0.072 & 1.08  & 0.40 &   8.572 & no  & \\ \hline

 7. & 1.4$\times$10$^{10}$ & 0.0450  & 3.10* & 0.00 &  5.913 & no  & 0.064 & 3.10* & 0.84 &  61.119 & yes &  Globular\\
    &                      &         &       &      &        &     & 0.068 & 1.00  & 0.40 &  13.223 & yes & \\ \hline

 8. & 1.3$\times$10$^{10}$ & 0.0200  & 3.10* & 0.00 & 34.012 & no  & 0.024 & 3.10* & 0.56 & 137.883 & yes & Uncertain \\
    &                      &         &       &      &        &     & 0.000 & 1.00  & 0.33 &  27.456 & yes & \\ \hline

 9. & 3.0$\times$10$^9$    & 0.0500  & 3.10* & 0.97 &  4.321 & yes & 0.042 & 3.10* & 1.26 &  12.802 & yes & Uncertain \\
    & 2.0$\times$10$^9$    & 0.0005* & 3.10* & 2.67 &  7.154 & no  & 0.061 & 2.63  & 1.11 &   9.039 & yes & \\
    & 2.0$\times$10$^9$    & 0.0005* & 3.10  & 2.67 &  7.154 & no  &       &       &      &         &     & \\ \hline

10. & 1.0$\times$10$^9$    & 0.0500  & 3.10* & 0.50 & 13.716 & yes & 0.042 & 3.10* & 0.15 &  54.800 & yes & Uncertain\\
    & 0.9$\times$10$^9$    & 0.0025* & 3.10* & 2.25 & 26.747 & no  & 0.000 & 1.00  & 0.15 &  30.674 & yes & \\
    & 0.9$\times$10$^9$    & 0.0025* & 2.69  & 2.06 & 17.238 & no  &       &       &      &         &     & \\ \hline

\end{tabular}
}
\end{center}
\end{table}

The preliminary tests reveal that in all cases $BI$-bands
dominate the values of $\chi_{min}^2$. These bands have the
largest systematics with respect to external photometry
(\cite[Massey et al. 2006]{Massey_etal06}). To account for that
and to minimize their impact we tentatively added 0.20\,mag to
the $B$-band and 0.12\,mag to the $I$-band magnitude errors. The errors listed in Table\,\ref{tab2} do not reflect this
modification. As a result, we have relatively more equal
contribution of the different colors to the $\chi_{min}^2$.

The globular cluster model fits much better the colors of most
targets than the elliptical model. The typical opacity $\tau_V$
across the M\,31 disk is $\sim$1\,mag. There are two exceptions
(objects No.\,8 and No.\,10) for which neither model yields a
reasonable match.

Interestingly, $R_V$ in M31 is lower than the typical Galactic
value of 3.1, and it is similar to the one obtained by
\cite[Savcheva \& Tassev (2002)]{SavchevaTassev02}. This may
indicate a smaller mean size of the dust grains in the diffuse
component of M\,31 ISM, in comparison with the Milky way. Although
the targets are located well within standard radius $R_{25}$, where
the active star formation still takes place, all of them are
projected in the inter-arm regions where the opacity of the disk
stays relatively law, as seen from the column density values in
Table\,\ref{tab1}. Here we assumed the Galactic gas-to-dust
ratio (\cite[Bohlin et al., 1978]{Bohlin_etal78}).

\begin{figure}[h]
\begin{center}
  \includegraphics[width=3.75in]{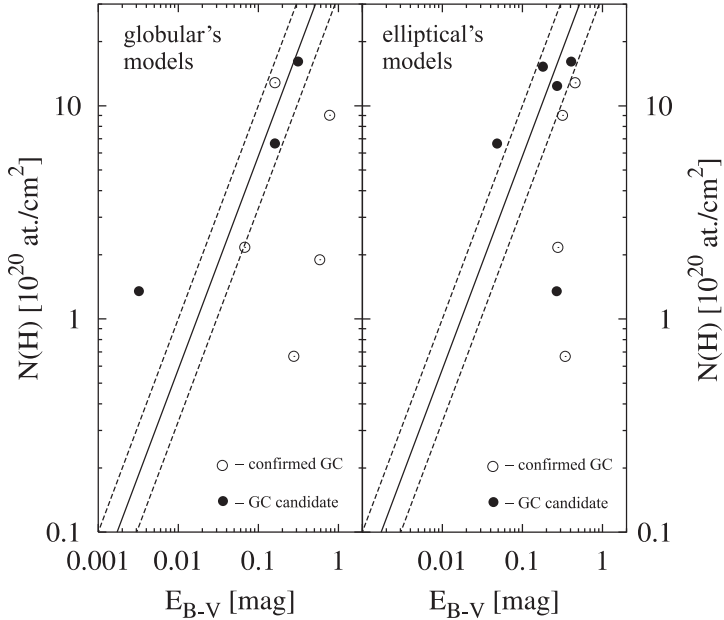}
\caption{The agreement between the total hydrogen column density
N(H) and the color excess E$_{B-V}$ with respect to
\cite[Kurth et al. (1999)]{Kurth_etal99} model colors of globulars
(left) and with respect to colors of ellipticals (right) as
predicted by the GALEV models (\cite[Bicker et al.
2004]{Bicker_etal04}). The extinction values are listed in
Table\,2 and correspond to $R_V$=3.1 and $\chi_{min}^2$. Dashed
lines represent the expected range and the thick line is the mean
Galactic gas-to-dust ratio (\cite[Bohlin et al.
1978]{Bohlin_etal78}).}
   \label{fig3}
\end{center}
\end{figure}

The relation between total gas column densities and the derived
extinctions, corresponding to $R_V$=3.1 and $\chi_{min}^2$, is
plotted in Fig.\,\ref{fig3}. Surprisingly, the derived extinctions
of the candidate globulars (targets No.\,6 and 9) correlate
better with the gas density if we use intrinsic colors derived
from the elliptical models. We contribute this to the spatial
variations of the reddening law.

Our analysis also considers the K-correction. We used the HyperZ
code of \cite[Bolzonella et al. (2000)]{Bolzonella_etal00}, that
includes a variety of spectral energy distributions for different
morphological types of galaxies. The results are presented in
Table\,\ref{tab4}. Both cases -- fixed to the elliptical type and
free morphological types yield reasonable $\chi_{min}^2$ values.
The extinction estimates are lower and the redshifts are higher
than those derived earlier, indicating a degeneracy between these
two quantities. The metallicity-opacity degeneracy is apparent in
Table\,\ref{tab3} as well: the higher is the abundance $Z$, the
lower is the derived extinction and vice versa. The HyperZ tends
to classify our targets as starburst galaxies, explaining the
larger extinction values in comparison with the case of fixed
elliptical morphological type. Note that a large fraction of the
extinction may be internal to a starburst galaxy and not related
to the M\,31 disk. This might be the case for targets No.\,8 and
10 which, together with No.\,9 are our best candidates for galaxies,
laying behind the M\,31 disk.

\begin{table}[t]
\begin{center}
\caption{Summary of the $\chi^2$ minimization considering the K-corrections,
for two different redshifts. The ``intrinsic'' redshifted colors of the galaxies
were determined with the HyperZ (\cite[Bolzonella et al. 2000]{Bolzonella_etal00}).
The numbers of the targets as the same as in Table\,1. The grid resolution is 0.05
along all axes and the Galactic extinction law of \cite[Allen (1976)]{Allen76} is
assumed. The rest of the columns are identical with those in Table\,\ref{tab3}.}
 \label{tab4}
 {\scriptsize
\begin{tabular}{|l||c|c|r|c|c||c|c|r|c|c|}\hline
{\bf No.} & {\bf Red-} & {\bf $A_V$} & {\bf $\chi_{min}^2$} & {\bf dust} & {\bf Galaxy} & {\bf Red-} & {\bf $A_V$} & {\bf $\chi_{min}^2$} & {\bf dust} & {\bf Galaxy} \\
          & {\bf shift}& {\bf [mag]} &                      & {\bf vs.}  & {\bf Type} & {\bf shift}& {\bf [mag]} &                      & {\bf vs.}  & {\bf Type} \\
          & {\bf $z$}  &             &                      & {\bf gas?} &           & {\bf $z$}  &             &                      & {\bf gas?} & \\ \hline

 1.       &      0.130 &        0.05 &               4.562  &      yes   &elliptical*&      0.135 &        0.05 &               0.673  &     yes    &starburst\\ \hline

 2.       &      0.200 &        0.55 &               1.743  &      yes   &elliptical*&      0.200 &        0.55 &               1.743  &     yes    &elliptical\\ \hline

 3.       &      0.145 &        0.70 &               1.109  &      yes   &elliptical*&      0.140 &        1.00 &               1.025  &     yes    &starburst\\ \hline

 4.       &      0.200 &        0.35 &               1.482  &      no:   &elliptical*&      0.150 &        1.45 &               1.365  &      no    &starburst\\ \hline

 5.       &      0.150 &        0.25 &               0.546  &      yes   &elliptical*&      0.150 &        0.75 &               0.464  &      no    &starburst\\ \hline

 6.       &      0.145 &        0.15 &               2.700  &      yes   &elliptical*&      0.155 &        0.05 &               1.807  &     yes    &starburst\\ \hline

 7.       &      0.145 &        0.15 &               3.926  &       no   &elliptical*&      0.350 &        0.10 &               1.051  &      no    &starburst\\ \hline

 8.       &      0.110 &        0.10 &               2.417  &       no   &elliptical*&      0.105 &        0.95 &               2.230  &     yes    &starburst\\ \hline

 9.       &      0.155 &        0.75 &               1.246  &      yes   &elliptical*&      0.155 &        0.75 &               1.246  &     yes    &elliptical\\ \hline

10.       &      0.050 &        0.00 &               4.659  &       no   &elliptical*&      0.150 &        0.50 &               2.571  &     yes    &starburst \\
\hline
\end{tabular}
}
\end{center}
\end{table}

\section{Conclusions}
We measure the opacity of the M\,31 disk from the color excesses
of 21 objects -- a mixture of galaxies behind the disk and globular
clusters. Seven of them are consistent with globulars, two - with
starburst galaxies and one - with an elliptical galaxy. Their
extinction estimates are consistent with a semi-transparent disk
($\tau_V\lesssim1$) in the inter-arm regions. We confirm the
conclusion of \cite[Savcheva \& Tassev (2002)]{SavchevaTassev00}
that the total-to-selective extinction value $R_V$ in the diffuse
ISM of M\,31 is on average lower than the typical Galactic value of
$R_V$=3.1. The gas-to-dust ratio appears similar to that in the
Milky way and it is independent from the galactocentric distance,
which might indicate Solar abundances along the line of sight
studied here (within 20$^{\prime}$$\div$85$^{\prime}$
from the M\,31 center).

\section*{Acknowledgments}

This work was partially supported by the following grants:
VU-NZ-01/06, VU-F-201/06 \& VU-F-205/06 of the Bulgarian Science
Foundation. One of the authors (P.N.) thanks to organizing
committee of IAU Symposium No.254 for the grant which allowed him
to participate.

\end{document}